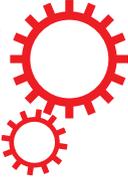

# SCIENTIFIC REPORTS

**OPEN**

# Angiogenic Factors produced by Hypoxic Cells are a leading driver of Anastomoses in Sprouting Angiogenesis–a computational study

Maurício Moreira-Soares[1], Rita Coimbra[2], Luís Rebelo[3], João Carvalho[1] & Rui D. M. Travasso[1]

Angiogenesis - the growth of new blood vessels from a pre-existing vasculature - is key in both physiological processes and on several pathological scenarios such as cancer progression or diabetic retinopathy. For the new vascular networks to be functional, it is required that the growing sprouts merge either with an existing functional mature vessel or with another growing sprout. This process is called anastomosis. We present a systematic 2D and 3D computational study of vessel growth in a tissue to address the capability of angiogenic factor gradients to drive anastomosis formation. We consider that these growth factors are produced only by tissue cells in hypoxia, i.e. until nearby vessels merge and become capable of carrying blood and irrigating their vicinity. We demonstrate that this increased production of angiogenic factors by hypoxic cells is able to promote vessel anastomoses events in both 2D and 3D. The simulations also verify that the morphology of these networks has an increased resilience toward variations in the endothelial cell's proliferation and chemotactic response. The distribution of tissue cells and the concentration of the growth factors they produce are the major factors in determining the final morphology of the network.

Sprouting angiogenesis–a process by which new blood vessels grow from existing ones–is an ubiquitous phenomenon in health and disease of higher organisms[1], playing a crucial role in organogenesis, wound healing[2], inflammation[3], as well as on the onset and progression of over 50 different diseases[4] such as cancer[5–7], rheumatoid arthritis[8] and diabetes[9]. This process leads to the formation of neo-vessel networks that irrigate tumors and other hypoxic tissues. In the circumstances in which these neo-vessel networks are functional, they present a ramified hierarchical tree-like structure at large scale, as observed in various tissues (from the mesenteric arteries, to the retina)[10–14], while at microvascular scale the capillaries form a lattice-like network connecting the arterial and venous vascular trees[10,14], as seen in reconstructed mouse vascular networks in ref.[15] and used in a tumor model in ref.[16]. To achieve this complex structure, the sprouting process involves different steps that are carefully regulated.

When tissue cells are in hypoxia, the Hypoxia Inducible Factor 1 $\alpha$ (HIF-1 $\alpha$) accumulates and is translocated to the cells' nucleus, where it works as a transcription factor for several genes encoding for proteins that take part in preparing the cell's response to hypoxia (see review[17] and also the mathematical models applied to inflammation and cancer in refs[18,19]). The Vascular Endothelial Growth Factor (VEGF) is one of these proteins and is necessary and sufficient to trigger angiogenesis. The VEGF secreted by hypoxic cells diffuses in the tissue, forming well defined spatial concentration gradients. VEGF is present in different isoforms with different binding capacity to the extracellular matrix (ECM) components[20,21] (the specific shape of the gradient depends also on this binding ability to ECM components)[22]. At a nearby vessel, VEGF promotes the loosening of the adhesion between vessel cells and triggers the activation of an endothelial cell (EC) that acquires the endothelial tip cell phenotype, as modeled in[23]. This cell leads a new sprout and migrates in the direction of increasing VEGF concentration, as observed in *in vivo* mice retinas in ref.[24]. The tip cell state promotes the activation of the endothelial stalk cell phenotype in the cells that follow its lead, as shown in ref.[25], a research paper on the mouse retina. Proliferation of the sprout stalk cells is triggered by both VEGF and the tension exerted by the tip cell, as predicted in a phase field model in ref.[26]. In the new growing sprout the cells behind the tip cell are able to take its place, acquire the tip cell phenotype for some tens of minutes, and drive sprout elongation, as displayed in studies *in vitro* and *in vivo* of tip-stalk cells dynamics in mouse retinal vessels[27,28]. This dynamic behavior ensures that there is always a cell at the front of the sprout with the tip phenotype capable of exerting a contractile force on the matrix, degrading and remodeling matrix fibres and opening a pathway for the sprout to grow.

[1]CFisUC, Department of Physics, University of Coimbra, Coimbra, Portugal. [2]Institute for Research in Biomedicine (iBiMED), Department of Medical Sciences, University of Aveiro, Aveiro, Portugal. [3]Department of Physics, Faculty of Sciences, University of Lisbon, Lisbon, Portugal. Correspondence and requests for materials should be addressed to M.M.-S. (email: mmsoares@uc.pt) or R.D.M.T. (email: ruit@uc.pt)

As the sprouts grow, ECs are able to either alter their shape or to extend and align vacuole in order to grow a lumen connected to the initial vessel that is capable of carrying blood[29,30], as shown in angiogenesis *in vitro* in ref.[31], *in vivo* in ref.[32] and used in modeling in ref.[33]. However, in order for the blood to flow, it is required a pressure gradient along the vessel, which is only achieved after the growing sprout merges either with an existing functional mature vessel or with another growing sprout that is in turn connected to an existing functional mature vessel. The process by which sprouts meet and merge is called anastomosis, studied in mouse models in[34,35] and also discussed in the review[36].

Though a pivotal step in angiogenesis, the mechanisms that drive vessel anastomosis are not yet completely understood. In particular, there are several unanswered questions regarding how a sprout is guided in the direction of other vessels in the tissue. Macrophages and microglial cells play an important role in this process *in vivo*, since they often sit at sites where vessels merge, stabilizing the connection between endothelial cells from different sprouts[34,37], demonstrated by *in vivo* experiments in the mouse retina in[38]. However, anastomoses still occur without the presence of macrophages or microglial cells, though less frequently[36,38].

Mechanical processes play an important role in driving vessel anastomosis. In fact, when a tip cell is next to an existing vessel, its filopodia are able to sense their microenvironment, and to establish contact with either other tip cells' filopodia or with other endothelial cells[24,38,39], as modeled in refs[36,40]. This contact is then strengthened by the VE-cadherins present in the filopodia[41], the tip cell advances in the direction of the existing vessel and it attaches to it[39]. The slow process of merging the two lumina at this new junction then ensues.

When the sprouts are at somewhat larger separations, mechanical factors are still able to play an important role in the formation of anastomoses. Take for example the situation of two tip cells meeting directly. Tip cells exert sufficient force to deform the matrix locally, and other tip cells are able to follow these deformation gradients[26,40,42]. In this way, a tip cell is able to sense the presence of other tip cells in its vicinity and will move in their direction. However, this mechanism still functions at rather short ranges, on the order of the cell size, as shown by *in vitro* experiments in ref.[43], and used in agent-based models in refs[42,44] only in the neighborhood of a tip cell the tissue is sufficiently deformed to guide nearby tip cells.

In some tissues other type of mechanical cues also drive anastomoses events at larger ranges. In the retina, for example, vessels grow above a layer of astrocytes that provide guidance and support[24]. In this situation, filopodia mediate the contact between the endothelial cells and the astrocytes network underneath, guiding the neo-vessels to almost replicate its structure. In this way, the astrocytes network plays an important role in driving vessel anastomoses in the retina.

However, in tissues where there is no underlying cell network for vessels to replicate, and where they explore the space between cells, we question how the vessels can be guided toward other vessels at larger scales in order to be close enough for mechanical communication to be relevant. In this work, we will demonstrate that pro-angiogenic factors, such as VEGF, play the decisive role in the process of anastomosis and in shaping the final network. There is strong experimental data that supports this hypothesis. In sprouting angiogenesis, these factors have a pivotal role in guiding the sprouts. Namely, VEGF in matrices is able to promote migration and anastomoses[24,45–47]. So much so that VEGF micro-patterns in 3D matrices are able to guide mouse endothelial cells *in vivo* to form anastomoses and functional vessels along those patterns[48]. Moreover, VEGF micro-beads also drive the formation of functional vessels in their vicinity *in vivo*[49,50].

In a non-pathological setting, hypoxic cells *in vivo* will be producing VEGF until they are irrigated, and they can only be irrigated after there are anastomoses in their surroundings. Therefore VEGF will keep being produced until nearby vessels merge and form loops capable of carrying blood. We will demonstrate that this mechanism can also account for an increased resilience of the network morphology towards endothelial cell proliferation and chemotactic response.

To address the capability of VEGF gradients to drive anastomoses in a neo-vessel network, we present an extensive and systematic 2D and 3D computational study of vessel formation in a tissue. The 2D model can be applied to angiogenesis in quasi two-dimensional vascular networks, such as networks in the retina, chick chorioallantoic membrane (CAM) assays, or other *in vivo* membranes. On the other hand, the 3D model can be applied to angiogenesis in 3D scaffolds and *in vivo* tissues, as the adipose tissue, and in tumor angiogenesis.

In the last two decades angiogenesis has been modelled using different approaches. Initially, researchers implemented continuous descriptions of angiogenesis that had the density of vessels as output[51–56]. These models were not aimed at obtaining the vascular network or the blood flow rate, but to provide an indication of the advancement of the irrigated region. Angiogenesis was also modelled using more detailed discrete agent-based models that take into account the interaction between every single cell in the system[57,58]. Discrete models have been coupled to tumor growth models, used to describe neo-vascularisation in the retina, and to study in detail the interaction between endothelial cells and their micro-environment[59–65]. Nevertheless, the large number of parameters and rules often difficult the full exploration of the parameter space in these models. Recently, hybrid models that combine a continuous description of the vessel sprout with a cell-based approach for tip cell creation and movement were developed[22,26,66]. These models allow the study of the shape of large vessel sprout networks with a smaller number of parameters. Nevertheless the role of chemical factors in regulating anastomoses formation in 2D and in 3D has not yet been studied. For reviews on the literature of angiogenesis modelling see for example[67–71].

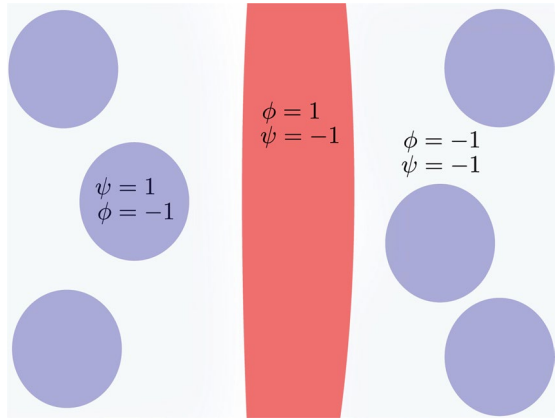

**Figure 1.** The value of the order parameters in different domains. The order parameter $\psi$ is $+1$ inside the tissue cells (represented in blue) and $-1$ outside, while the order parameter $\phi$ is $+1$ inside the endothelial cells (represented in red) and $-1$ in the rest of the domain. The stroma is characterized by the region where $\phi$ and $\psi$ are both negative. The model dynamics prevents the superposition between the positive domains of the two order parameters.

In the next section we will introduce the computational model used. In the Results and Discussion section we will present and compare the results of the 2D and 3D models regarding the capability of VEGF to drive anastomoses formation. It will be stressed the role of this mechanism in determining the morphology of the vascular networks, rendering it less dependent on endothelial stalk cell proliferation rates or on the migration velocity of tip cells. Finally we draw the conclusions of the work in the last section.

## Methods

The angiogenesis mathematical model implemented is a natural extension of the model introduced in[22], and further explored in[40,72–74]. Here, we will review, in an abbreviated way, this mathematical model underlining the two major improvements that are introduced: the description of the cells in hypoxia using a second order parameter, and the estimation of tissue irrigation to couple it with hypoxia regulation.

We model the ECs and the ECM through a phase-field formalism[22,75,76], defining an order parameter, $\phi(\mathbf{r}, t)$ that identifies the vessel network: the value for this order parameter is approximately equal to 1 inside the vessel network, and equal to $-1$ outside. In the present work, we introduce a second order parameter to describe the tissue cells $\psi(\mathbf{r}, t)$, which is equal to 1 inside these cells and $-1$ outside. In this way, the vessels, ECM and tissue cells can be unequivocally identified by the local values of $\phi$ and $\psi$ (see Fig. 1).

The equation for the evolution of the order parameter $\phi(\mathbf{r}, t)$ is composed of two main terms: one describing the dynamics of the interface, and another describing the proliferation of endothelial cells (function of the angiogenic factor concentration, $T$):

$$\partial_t \phi = \underbrace{M\nabla^2[\phi^3 - \phi - \varepsilon^2\nabla^2\phi + \gamma(\phi^2 - 1)(\psi - 2)(\psi + 1)^2]}_{\text{Interface Dynamics}} + \underbrace{\alpha_p(T)\,\phi\,\Theta(\phi)}_{\text{EC Proliferation}}. \quad (1)$$

The first term is directly obtained from the system's phenomenological free-energy functional (see Supplementary Information S1), which includes a surface tension term[22] and an energy cost for the overlap of vessels and tissue cells[77]. The $\gamma$ in (1) is proportional to this energy cost and is set high enough in order to avoid appreciable overlap in our simulations. In equation (1) $M$ is the mobility coefficient, $\varepsilon$ is the width of the capillary wall (proportional to the surface tension), and $\Theta(\phi)$ is the Heaviside step function. The proliferation rate is represented by the function $\alpha_p(T) = \beta_p T$, for $T < T_p$, with $\beta_p$ constant; $T$ being the angiogenic factor concentration, and $T_p$ the angiogenic factor concentration for which the proliferation reaches its maximum value. For $T > T_p$ we set $\alpha_p(T)$ to the constant value $\alpha_p(T) = \beta_p T_p$. Due to the Heaviside function, the proliferation only occurs inside the capillary.

Often hypoxic cells have a low motility, especially if they are part of an epithelial tissue. Therefore, we consider in the simulation that tissue cells are static and do not change their shape. Hence, the order parameter $\psi$ will not be altered during simulation. Nevertheless, the simulation methods used permit to introduce motility and deformation on the hypoxic cells, which will be explored in future work.

The angiogenic factor is modeled as a diffusive field $T$, that is consumed by the ECs with the rate $\alpha_T$:

$$\partial_t T = D\nabla^2 T - \alpha_T \phi T \Theta(\phi). \quad (2)$$

The concentration of angiogenic factor is kept at a constant value $T_s$ at the center of all active hypoxic cells. During the simulation, the sources will be deactivated as the vessel network grows and as they are irrigated.

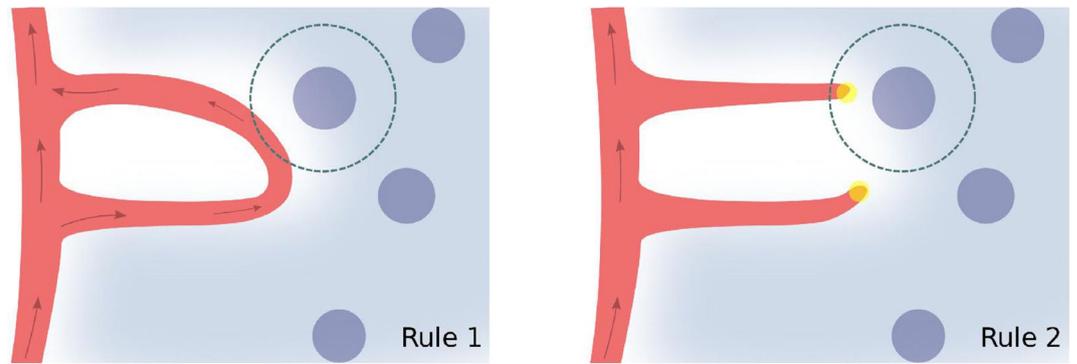

**Figure 2.** Schematic representation of the two different rules for hypoxia regulation. In the simulation with Rule 1 a cell in hypoxia will deactivate its production of VEGF when the blood circulation reaches a site at a distance from the cell shorter than the oxygen diffusion length. In the model with Rule 2, the VEGF production will cease if any vessel reaches a point closer to the cell than the oxygen diffusion length, independently of whether there is blood circulating in that vessel or not. The endothelial cells are represented in red, the tip cells are the yellow circles and the angiogenic factor is represented in light blue. The cells in hypoxia are represented as darker blue circles. The arrows indicate the vessels with blood flow.

The endothelial tip cells are represented by agents with radius $R_c$. A new agent is introduced whenever there is a minimum angiogenic factor concentration $T_c$, a minimum angiogenic factor gradient $G_m$, and if the tip cell candidate is located at a distance larger than $4R_c$ from all other tip cells. In this way, we are able to retrieve the typical "salt and pepper" pattern of tip/stalk cells arising from the Delta-Notch signaling[23,27,39]. Some recent works[78] have modeled the Delta-Notch signaling pathway in the context of angiogenesis also taking into account the Jagged1 ligand, which contributes to the stabilization of the ECs' phenotypes[78,79], possibly influencing the tip/stalk cell pattern. The implementation of this more complex Delta-Jagged-Notch mechanism is out of the scope of the present work but it is expected to be introduced in future model developments.

All the activated tip cells migrate by chemotaxis, and their velocity is aligned with the local angiogenic factor gradient. The tip cell velocity is given by $\mathbf{v} = \chi \nabla T$, for $|\nabla T| < G_M$, with $\chi$ constant, $G_M$ being the gradient modulus for which the maximum velocity is attained. For $|\nabla T| \geq G_M$ the tip cell velocity reaches its maximum value $\chi G_M \nabla T / |\nabla T|$. A tip cell is not able to enter inside a tissue cell (independently if the cell is in hypoxia or not): when the distance between the tip cell and a tissue cell is lower than $2R_c$, the radial component of the velocity is set to zero and the tip cell moves in the azimuthal direction around the tissue cell.

The value of $\phi$ inside the tip cell is set to a constant value that depends on the tip cell velocity and on the local ECs' proliferation rate[22].

Blood is able to flow once there are anastomoses in the vascular network. In order to estimate the irrigated regions in the tissue, we first identify the vessels where the blood is flowing. We start by extracting the medial lines of the vascular structure, thus finding the network bifurcation sites and the length of every vessel. The extraction method used for this purpose is detailed in[80]. Afterwards, we calculate the blood flow rate in every vessel (see Supplementary Information S1). In the simulation there is a single input node (at the top of the simulation box) and a single exit node (at the bottom of the simulation box). For more details regarding the mathematical model and for the values of all parameters please refer to the Supplementary Information S1.

**Cells in Hypoxia - Rules 1 and 2.** With the aim of addressing the capability of VEGF to drive anastomosis formation, we compare the results of the model obtained with two alternative rules concerning the tissue angiogenic factor production. When the new vessels are capable of delivering oxygen to the tissue, the newly irrigated cells deactivate their hypoxia mechanism and cease the production of angiogenic factors. Therefore several models in the literature[10,14,60,81] use estimates of tissue irrigation in order to determine which cells are producing angiogenic factors. To mimic this mechanism we consider a simplified rule to identify the VEGF producing cells:

**Rule 1:** A cell produces angiogenic factor in the model if it is located at a distance larger than the average oxygen diffusion length ($d = 25\,\mu$m)[82] from all vessels with blood flow rate different from zero (Fig. 2 left).

This is the simplest rule guaranteeing that anastomoses events are required for hypoxic cell deactivation. However, many other models focus on the ability of the angiogenic factor to drive vessel growth and not anastomoses[22,40,83,84]. These models implement a different rule for tissue cell deactivation:

**Rule 2:** A cell produces angiogenic factor in the model if it is located at a distance larger than the average oxygen diffusion length ($d = 25\,\mu$m) from all vessels (Fig. 2 right).

In short, with Rule 2 chemical guidance promotes sprouting but is not associated with tissue irrigation, while in Rule 1 it is. Comparing the morphology of the networks formed under these two rules, we show how important it is for tissue irrigation to regulate VEGF production to create functional vascular networks in 2D and in 3D.

The calculation of tissue irrigation is in general quite complex, being dependent on the oxygen pressure at every vessel, on vessel diameter, flow rate (function of the non-constant viscosity due to variable hematocrit and the Fahrëus-Lindqvist effect), hematocrit level (which is altered at non-symmetrical bifurcations), oxygen diffusion constant in the tissue and oxygen uptake by tissue cells[85–87]. However, in the present work a detailed calculation of tissue irrigation will not be required, since our aim is to demonstrate that the need for anastomoses formation to stop VEGF production in tissues (as in Rule 1) will determine several characteristics of the observed vascular morphology. Rule 1 keeps the model general, with all anastomoses providing the same amount of irrigation to the tissue. With a full calculation of tissue irrigation we would observe that the irrigation provided by each anastomosis would depend on its location within the vessel network. This systematic study of vessel density in growing vasculatures as a function of the geometry of the initial network and of the oxygen pressure is very relevant, but it is outside the scope of the present article.

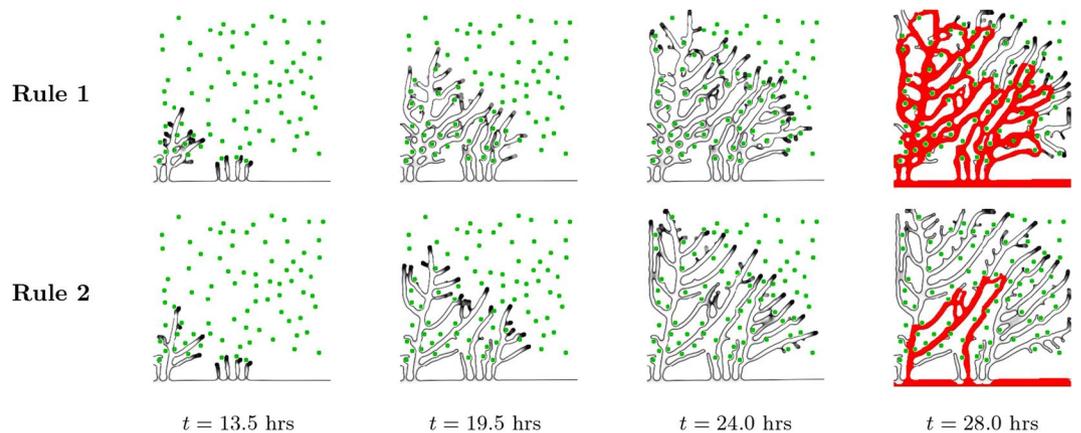

**Figure 3.** Vessel growth in two dimensions. Evolution of a vascular system with an initial capillary at the bottom and a random distribution of hypoxic cells (green circles), which drive the chemotactic response by ECs. Four time points are represented. The first row of figures correspond to the vessel network grown under the Rule 1 for hypoxic cell deactivation, while the second row of images correspond to the vessel network grown under Rule 2. On the final column the irrigated vessels are identified in red. The number of anastomoses and the number of vessels with circulating blood are much higher when Rule 1 is implemented.

In this way, the final vasculatures obtained by running the simulation with Rule 1 will be directly compared to the simulations where the angiogenic factor drives growth but not anastomoses (using Rule 2). We will then draw conclusions regarding the capacity of VEGF gradients to drive vessel anastomoses.

## Results and Discussion

**Anastomoses in 2D.** Similarly to the setting in[22] we start with a random distribution of hypoxic cells in a simulation box of size $375 \times 375\,\mu m^2$. These hypoxic cells produce the angiogenic factor which diffuses slowly in the tissue. When the angiogenic factor concentration at the vessel initially located at one edge of the box is larger than $T_c$, a new tip cell is introduced in the simulation. This tip cell moves in the direction of the local angiogenic factor gradient forming a new sprout. Progressively, new tip cells are activated and a ramified network is produced (see Fig. 3). As the tip cells migrate in the tissue they are able to contour the hypoxic cells, diverting their movement direction as they approach these cells.

The morphology of the obtained networks is strikingly dependent on the rule for hypoxic cell deactivation. With Rule 2, the hypoxic cells start deactivating once the vessels pass nearby, leading to an open network with fewer vessels and anastomoses. The vessels are straighter and directed towards the hypoxic tissue. As a result, only few vessels are able to carry circulating blood. On the other hand, with Rule 1, the number of branches is higher. The vessels surround several cells in hypoxia, irrigating them. The resulting vessel network morphology is more lattice-like instead of tree-like. Many more vessels carry blood flow. The observed vasculature also suggests that with Rule 1 regions with more hypoxic cells have a higher density of vessels, while regions with less cells have a lower vessel density.

To verify how prevalent these differences between vessel morphology are, we carried out a systematic quantitative study of the morphology obtained from our simulations in 2D and 3D. To characterize the vascular trees we measure their branch density, average vessel diameter, and density of anastomoses. These descriptors were evaluated as a function of the two model parameters that were considered most relevant for the network evolution[22,88]: the maximum velocity of tip cell migration in the tissue (equal to $\chi G_M$ in the model, though only the value of $\chi$ is altered), and the highest stalk cell proliferation rate (equal to $\beta_p T_p$ in the model, only the value of $\beta_p$ is altered). Notice that the proliferation term in the model corresponds to the increase of vessel area (or volume, in 3D) per unit time: it is a measure of the change of space occupied by the vessel per unit time[22]. This description is ideal to be directly compared with observations of the network evolution through diverse imaging techniques. However, since the vessels are hollow, the cell proliferation *in vivo* is expected to be systematically smaller than the cell proliferation (space increase) term in the model. Nevertheless, since the neo-vessels resulting from sprouting have all similar calibers (see for example[66] for *in vivo* results) and since also in our simulation the neo-vessels have all approximately the width, the extra proliferation in our simulation serve simply to fill-in the lumen. Therefore, we expect that, even taking into account the vessels lumen, the progress of the network morphology as a function of the cells' proliferation rate would be the same as the one observed in our simulations.

The results of this analysis are presented in Fig. 4. Each data point in these plots represents the average value from three random distributions of hypoxic cells. We start from a random distribution of hypoxic cells to avoid any spatial bias. For both simulations with Rules 1 and 2, as the proliferation rate is increased, the vascular networks become more complex, with a larger number of vessels and anastomoses. With Rule 2, for very low proliferation rates, the number of vessels is very small and there are no anastomoses (see Fig. 4A,B). As the proliferation rate increases, the number of vessel branches increase rapidly until it reaches a high stationary number of about 700 branches per mm$^2$. This increase in vessel branch density is accompanied by an increase in anastomoses density. The average vessel diameter increases only slightly with the stalk cell proliferation rate (Fig. 4C).

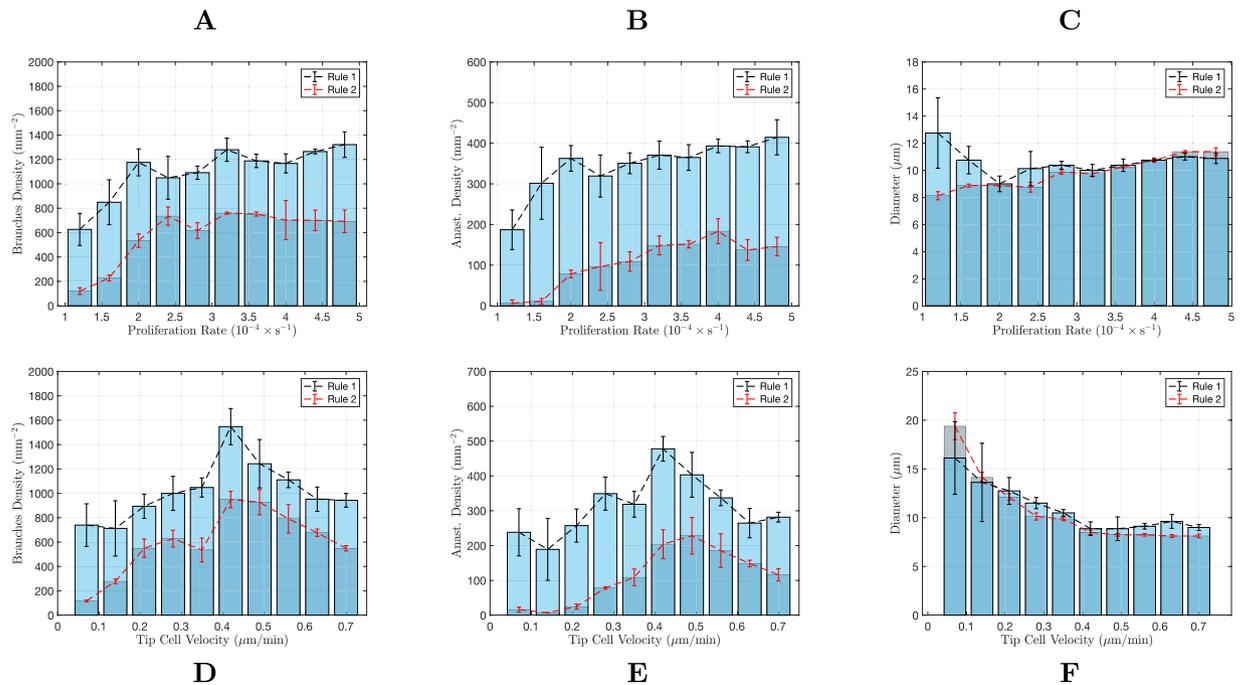

**Figure 4.** Characterisation of the vascular networks in two dimensions. Comparison of the branches density (**A**,**D**), anastomoses density (**B**,**E**) and average vessels diameter (**C**,**F**), between the two rules implemented in the model, as a function of the maximum proliferation rate (top row) and as a function of the maximum tip cell velocity (bottom row). The networks were analysed after they had grown sufficiently to deactivate the VEGF production in all tissue cells. The error bars denote the error in the average of 3 simulations (for each condition).

A similar qualitative behaviour regarding the number of vessel branches and anastomoses is observed with Rule 1. However the values of these two descriptors are considerably higher. Remarkably, the vasculature morphology is much more stable. In the range of proliferation explored in the simulation, we obtain an increase of approximately ten-fold in the number of branches and anastomoses when using Rule 2 for hypoxic cell deactivation, whereas in the simulation with Rule 1, this increase was only about two-fold. For most of the range of proliferation rates explored, the average vessel diameter did not vary when Rule 1 was implemented.

Similar results were observed for variable tip cell velocity, namely, Rule 1 leads to a clear increase in the number of branches and anastomoses, but at the same time, to more stable vascular networks (Fig. 4D–F). For both rules for hypoxic cell deactivation, as the tip cell velocity increases, the number of branches increases until a maximum is reached. After this maximum, the velocity of the tip cell is so high that the sprouts are initially not thick enough to accommodate tip cell activation (the vessels start being thinner than two cell radii). The activation of new tip cells can then only happen after there is enough proliferation in these thin vessels. Because of the increased difficulty in tip cell activation, the number of branches decreases for high tip cell velocity. The density of anastomoses in the network follows, qualitatively, the non-monotonic behaviour of the branch density. For Rule 1 the number of branches and anastomoses are significantly higher and their relative variations with tip cell velocity are significantly smaller (for example, with Rule 1 the number of anastomoses varies only about 2-fold within the range of tip cells velocities explored, while with Rule 2 it varies about 15-fold within the same range). With respect to vessel diameter, we observe that the vessels become thinner with increasing tip cell velocity until they reach a constant value for tip cell velocities larger than the velocity where the maximum branch density is attained. Figure 4F suggests a similar change in vessel diameter with tip cell velocity with both Rules 1 and 2.

We conclude that, in two dimensions, the hypoxic cell VEGF production levels can force anastomoses to be created in the system. This mechanism also renders the morphology of the vascular networks much more insensitive to the proliferation rate and tip cell velocity. The hypoxic cells are the main factor in determining the structure of the vasculature that will satisfy their needs.

These conclusions are robust to the inclusion in the model of more realistic relations of, for example, the chemotactic response on the degree of hypoxia in the cells. To explore the effect of variable chemotactic response, we introduce a simplified dependence of $\chi$ in $T$, inspired in the experimental evidence that the expression levels of VEGF receptor 2 in ECs are lower in hypoxia[89,90]. Hence, we considered a lower chemotactic response in this case. We repeated the analysis reported in Fig. 4 with the chemotactic response as a function of the hypoxia level (we used the local VEGF concentration, $T$, as proxy for hypoxia, $\chi = \chi_0(1 - T)$), and we did not observe any significant change in the observed trends (see Supplementary Information S1).

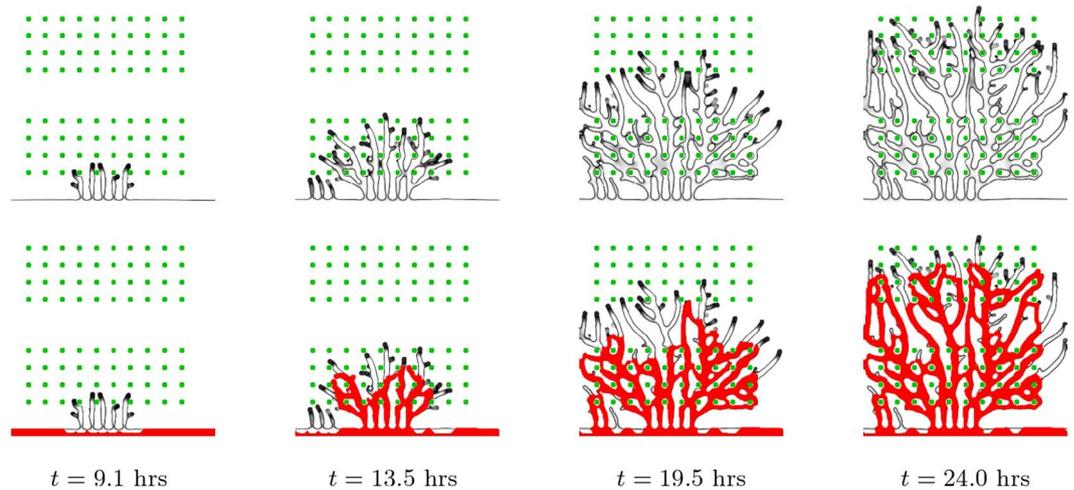

**Figure 5.** Hypoxic cell distribution determine vascular network. Evolution of a vascular system with an initial capillary at the bottom and uniformly distributed hypoxic cells (green dots), sources of VEGF which drive the chemotactic response by ECs (maximum proliferation $2.8 \times 10^{-4}$ s$^{-1}$ and maximum tip cell velocity 0.375 $\mu$m/min). Four time points are represented. The bottom row shows the same network growth with the vessels with blood flow in red.

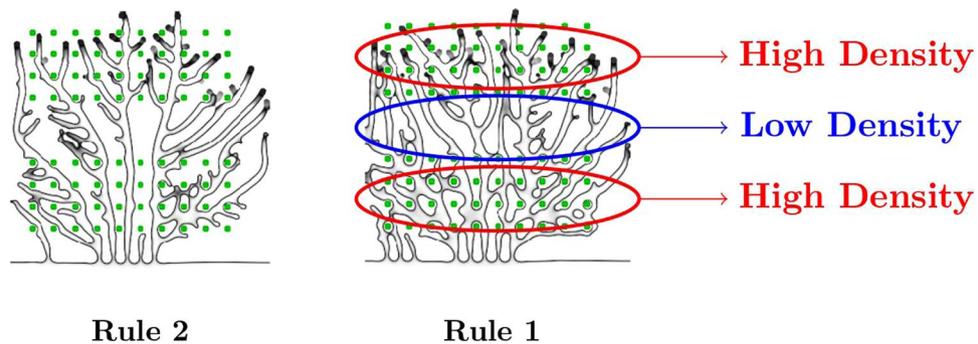

**Figure 6.** Neo-vessel morphology after 24 hours using the two rules. In the right panel (Rule 1) we mark the regions of high vessel density, on the top and bottom, where the hypoxic cells are placed, in contrast with the lower vessel density in a middle region (without hypoxic cells). With Rule 2 (left panel) the vessel density is much more uniform throughout the domain.

To address the capability of hypoxic cell distribution to determine vascular structure, we run the two dimensional model with Rule 1 for a regular distribution of hypoxic cells (see Fig. 5). In this simulation we set eight parallel rows of ten hypoxic cells with a large space between rows 4 and 5. The evolution of the system clearly shows how the neo-vascular network grows and branches in order to feed the first rows of hypoxic cells. The branching then almost stops when the capillaries cross the region without VEGF sources, going back to a very branched morphology when they reach the second domain with hypoxic cells.

A late instant of the growth of this network is again depicted in Fig. 6 right, where it is evident that the density of new vessels is higher, with more branches and anastomoses in the hypoxic regions, on the top and on the bottom of the figure. On the region without cells in hypoxia (middle area) the branch density is lower and the number of anastomoses is reduced (for example, the branch density in this middle region is $0.81 \times 10^3$ mm$^{-2}$; compare to a branch density $1.8 \times 10^3$ mm$^{-2}$ on the top hypoxic region of the system). Strikingly, in the corresponding network obtained from the simulation with Rule 2 (see Fig. 6 left), the hypoxic cells do not affect dramatically the vessel density, which is approximately constant throughout the simulated domain (with Rule 2 the branch density in the middle region is $0.97 \times 10^3$ mm$^{-2}$, similar to a branch density $1.1 \times 10^3$ mm$^{-2}$ on the top hypoxic region).

Remarkably, using solely the production of chemical factors, and without including mechanical components, we are able to guide the vasculature to form anastomoses where they are required. Similarly, sources of VEGF *in vivo* are able to promote the anastomosis and formation of new vessels in their vicinity[49,50]. This result clearly agrees with what is expected in angiogenesis due to hypoxia: a higher density of new vessels in the region with more hypoxic cells. The vessel network is tailored to feed the tissue.

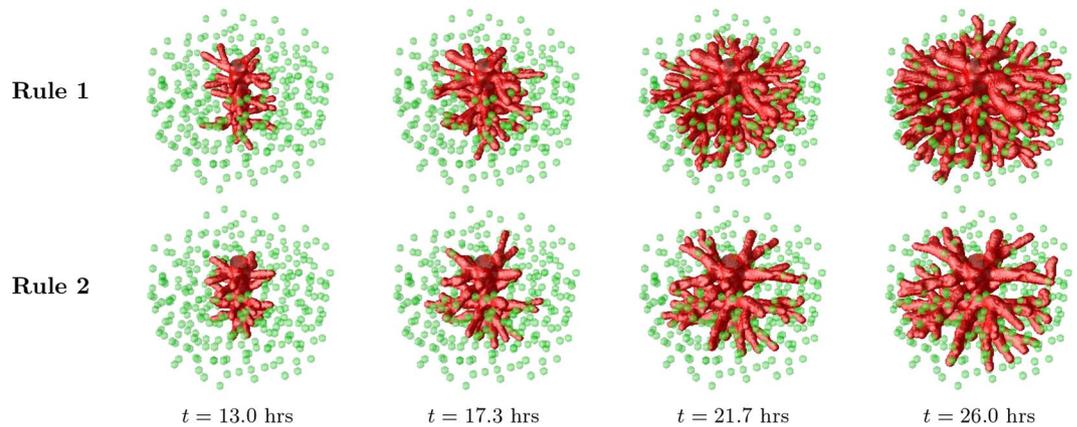

**Figure 7.** Vessel growth in three dimensions. Evolution of a vascular system with an initial capillary at the center and randomly distributed hypoxic cells (green spheres), sources of VEGF, which drive the ECs chemotactic response. Four time points are represented. The first row of images corresponds to the results for the model under the Rule 1, while the bottom row to Rule 2. We clearly observe a higher density of vessels in the simulation with Rule 1.

We conclude that, in two dimensions, the distribution of hypoxic cells and the production rate of angiogenic factors are the main regulators of angiogenesis, not the proliferation of endothelial cells or the tip cell velocity. This conclusion raises questions regarding the power to change the vessel network by the newly proposed anti-angiogenic and vessel normalization therapies that target the metabolism of endothelial cells in a way to control their proliferation or migration velocity[91–94]. With these therapies, and without targeting VEGF producing cells, only when these parameters are dramatically changed, the network morphology is critically altered (two caveats: first, with lower tip cell velocity the vasculature would grow at a slower rate, but the number of branches and anastomoses, when comparing networks of the same size, would not be significantly altered; second: the simulation is run with the assumption that hypoxic cells greatly decrease their VEGF production rate once they are irrigated, which does not happen, for example, in several high grade malignant tumors)[4]. However, the change of endothelial cell proliferation or migration rates may have an important effect in the quality of the vessel (e.g. lumen size, leakage, maturation)[93].

However, vessels in most cases grow in three dimensional networks. Therefore, we extend the simulations to 3D to verify to which extent the conclusions regarding VEGF driven anastomoses hold in the formation of 3D vascular networks.

**Anastomoses in 3D.** In this subsection we are looking to verify if, like in two dimensions, the production of angiogenic factors by the hypoxic cells is able, by itself, to guide vessels to meet each other in three dimensions. We are also interested in understanding if the characteristics of endothelial cells, such as their proliferation rate and tip cell velocity, have an important role in shaping these networks in three dimensions; or if, like in two dimensions, modifications in these properties lead only to moderate changes in the vascular network.

The neo-vasculature in this three dimensional model is simulated in a box of size $250 \times 250 \times 125\ \mu m^3$, it starts from a central capillary[72], and the new sprouts grow along the gradient of angiogenic factor, produced by hypoxic cells distributed randomly in the matrix (see Fig. 7). As in the two dimensional model, a vessel is able to contour a hypoxic cell it approaches.

As in two dimensions, the vessel network grown with Rule 1 is clearly more dense than the vessel network obtained from the simulation with Rule 2 (see Fig. 7). At the end of the simulation with Rule 1 all the hypoxic cells were irrigated, a consequence of the large number of anastomoses present in the network.

A systematic study of the density of branches, anastomoses and vessel diameter of the simulated three dimensional networks was also carried out (see Fig. 8). Each data point in these plots represents the average value from three random distributions of hypoxic cells.

Strikingly, anastomoses are almost absent in the simulations with Rule 2, independently of the proliferation rate and tip cell velocity. This is different from what was observed in 2D, where Rule 2 was able to lead to the formation of a large number of anastomoses for certain values of the parameters. In three dimensions, Rule 1 leads to an increase in the branch density and to an impressive ten-fold increase in the density of anastomoses in the network. Hence the production of VEGF until the tissue is irrigated is able by itself to guide the growing sprouts toward each other, in order to form new anastomoses.

We also observe differences with the two dimensional scenario regarding the dependence of the network morphology on the proliferation rate. Similarly to the two dimensional simulation, for larger endothelial cell proliferation rate, the number of branches increases. However, in 3D this increase is smoother and spans a wider range of proliferation rates. The increase on the number of branches with the proliferation rate occurs both in the simulations with Rules 1 and 2. Regarding the density of anastomoses, we observe a plateau for higher proliferation rates. However, neither the number of branches nor the number of anastomoses can be said to be almost independent on the proliferation rate, as it occurred in 2D.

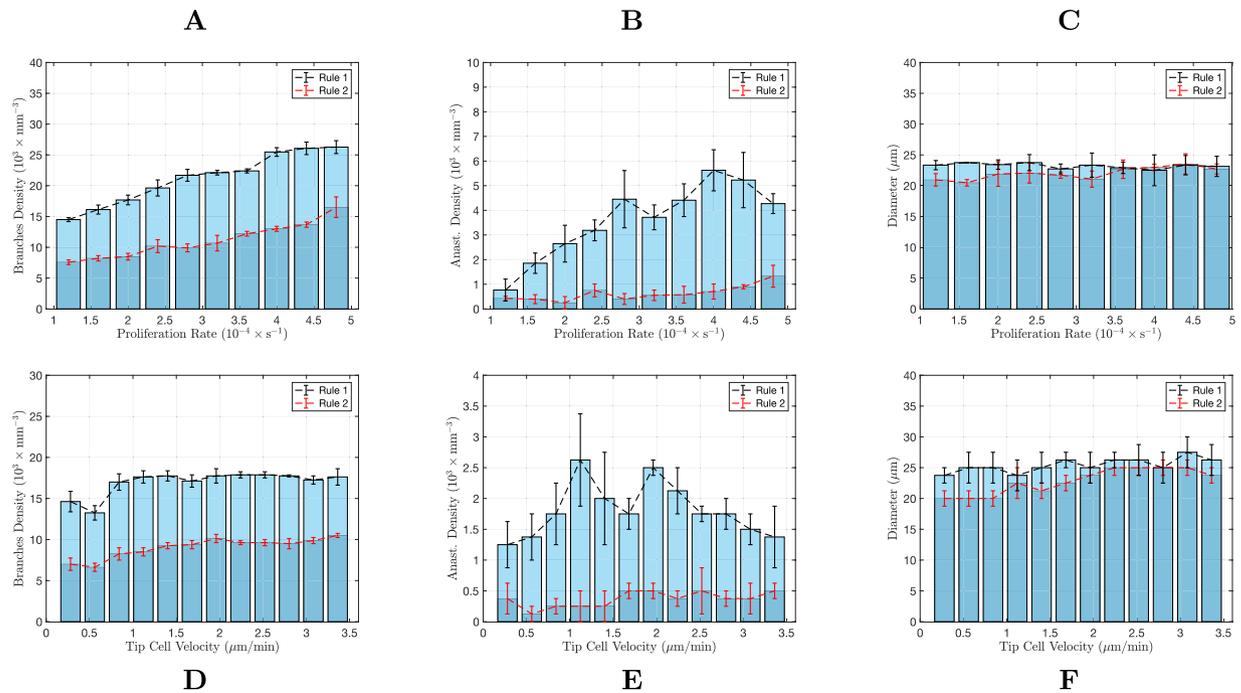

**Figure 8.** Characterisation of the vascular networks in three dimensions. Comparison of the branches density (**A**,**D**), anastomoses density (**B**,**E**) and average vessels diameter (**C**,**F**), between the two rules implemented on the model, as a function of the maximum proliferation rate (top row) and as a function of the maximum tip cell velocity (bottom row). The networks were analysed after they had grown sufficiently to deactivate the VEGF production in all tissue cells. The error bars denote the error in the average of 3 simulations (for each condition).

On the other hand, when varying tip cell velocity, the relative change in network morphology parameters is much smaller for both Rules 1 and 2. Though for lower tip cell velocities the vessel network grows slower, and thus takes a longer time to reach the stationary state at the end of the simulation their final morphology is rather independent of the value of the maximum tip cell velocity value.

The diameter of the vessels is not altered considerably neither with the increase in proliferation nor in tip cell velocity. With the increase in proliferation there are simply more vessels, rather than thicker vessels. In the case of increasing tip cell velocity, the network is rather stable, with the number of branches, anastomoses and branch diameter not varying appreciably.

## Conclusion

In this article we presented a model of vessel sprouting that addresses the VEGF gradients' role in driving the formation of anastomoses. This important biological mechanism is fundamental to irrigate tissue cells. However, there are still several unanswered questions regarding how a sprout is guided in the direction of other vessels in the tissue. The role of mechanical cues in anastomose formation has been explored in the literature[24,26,36,38–40,42,43] but it can only account for guidance at short distances between merging sprouts. In the present work we demonstrated how slowly diffusible growth factors gradients, maintained by tissue cells until they have access to sufficient oxygen and nutrients, are able to guide vessel sprouts toward each other at larger scales. This mechanism is particularly relevant in three dimensions, where without it the network has a reduced number of anastomoses, and it is not capable of irrigating the tissue.

The role of hypoxic cells is therefore pivotal in determining the final morphology of the network. We showed that it is the location of these cells and the concentration of the growth factors they produce that better determine the number of vessels and of anastomoses in the final network. Actually, these factors are more relevant to the final morphology than the tip cell chemotactic response or the ECs' proliferation rate. Experimentally it was shown the extraordinary capacity for growth factors spatial distributions *in vivo* to influence sprouting[48,95] and to determine the site of functional neo-vessels[49].

This resilience of the final network morphology toward EC's properties in 2D and in 3D (to a smaller degree) guarantees reliable vascular structures in non-pathological settings. However, it raises questions regarding the capacity for modifying the vessel network of the newly proposed vessel normalization therapies, which control EC's metabolism[91,92] in tissues where cells are still able to regulate their VEGF production. With these therapies, and without targeting VEGF producing cells, the network morphology is only critically altered when the EC's proliferation rate or migration velocity are dramatically affected. In fact, in agreement with our work, it has been observed experimentally *in vivo* that a reduction in the EC's proliferation rate of over 50% does not significantly alterthe density of tumor vessel networks[93]. Moreover, to observe a dramatic change of the vasculature, one has to produce a substantial change in EC's metabolism[96], leading to very low proliferation rate and migration velocity, just as predicted by our work. However, a moderate decrease in EC's proliferation is able to affect the capacity of the network to irrigate the tissue due to a lower vessel leakage and an increase in lumen size[91–94]. However, in conditions where tissue cells produce VEGF at permanently high rates independently of the irrigation provided by the network, both the ECs' proliferation rate and the tip cell velocity are able to shape the vascular structure[88].

In this mathematical model, the hypoxic cells function as sites for anastomoses formation, producing angiogenic factors until vessels merge. This suggests that *in vivo* cells that promote anastomosis, such as macrophages and microglia[34,37,38] can fulfil this role by chemical guidance (from VEGF and other chemical factors). It is also evident the relevance of targeting macrophage metabolism in anti-angiogenic therapy[97].

### Acknowledgements

The authors thank fruitful discussions with Lino Ferreira and Paulo Matafome. The simulations were done in the Navigator cluster of the University of Coimbra. This work was funded by FEDER funds through the Operational Programme Competitiveness Factors - COMPETE and by national funds by FCT - Foundation for Science and Technology under the strategic project UID/FIS/04564/2016 (M.M.S., J.C., R.D.M.T.). M.M.S. acknowledges the support of the National Council of Technological and Scientific Development (CNPq) under the grant 235101/2014-1. RT acknowledges FCT's support through the FCT Researcher Program.

### Author Contributions

M.M.-S., J.C. and R.D.M.T. conceived the study; M.M.-S. and R.D.M.T. wrote the code; M.M.-S., R.C. and L.R. run the simulations; M.M.-S., J.C. and R.D.M.T. wrote the manuscript.

### Additional Information

**Supplementary information** accompanies this paper at https://doi.org/10.1038/s41598-018-27034-8.

**Competing Interests:** The authors declare no competing interests.

**Publisher's note:** Springer Nature remains neutral with regard to jurisdictional claims in published maps and institutional affiliations.